\documentstyle[amssymb,sprocl]{article}
\input{psfig}
\def\Journal#1#2#3#4{{#1} {\bf #2}, #3 (#4)}

\def\PRL{\em Phys. Rev. Lett.}
\def\PRB{{\em Phys. Rev.} B}

\begin{document}

\title{RANDOM TILING TRANSITION IN THREE DIMENSIONS}

\author{W. EBINGER, J. ROTH, H.-R. TREBIN}

\address{Institut f\"ur Theoretische und Angewandte Physik\\
Universit\"at Stuttgart, D-70550 Stuttgart, Germany}

\maketitle\abstracts{
Three-dimensional icosahedral random tilings are studied in the
semi-entropic model. We introduce a global energy 
measure defined by the variance of the quasilattice points in
orthogonal space. The specific heat shows a pronounced \textsc{Schottky} type
anomaly, but it does not diverge with sample size. 
%which may indicate a phase transition from an ordered 
%quasicrystal to a random tiling.
The flip susceptibility as defined by Dotera and
Steinhardt [Phys.~Rev.~Lett.\ {\bf 72}, 1670 (1994)] diverges and shifts to lower temperatures, thus indicating a transition at $T=0$. 
Contrary to the
Kalugin-Katz conjecture, the self-diffusion
shows a plateau at intermediate temperature ranges which is
explained by energy barriers and a changing number of flipable
configurations.}

\section{Introduction}

The stability of quasicrystals has been a subject of intensive
research since they were discovered in 1984. In the random tiling
(rt) model 
stability is ascribed to the entropy $S$. If the internal energy $U$ also
contributes one is dealing with the semi-entropic model characterized
by a free energy $F(T)=U(T)-TS(T)$.
The rt model is an  
abstraction where tiles cover the whole space
without gaps. Thermal fluctuations and deformations of the tiles and
phonon degrees of freedom are neglected. The only dynamic motion that
exists is a local rearrangement of tiles, called ``flips''. 

\section{Random tiling characterization} % 6.
\label{character}\label{variance}

A random tiling may be characterized by the mean square deviation
of the point distribution from the center of mass in the orthogonal
space. The variance is defined by $\Omega=\overline{{\bf y}^2} -
\overline{\bf y}^2$~(1), where the average is over all vertices $N$, 
and ${\bf y}_i$ are the position
vectors in orthogonal space.
The lattice points of a quasicrystal are distinguished by their
local environments. There are 24 allowed vertices, but 5450
vertices may occur in a random tiling. 
In the icosahedral rhombohedron tiling some 
flips change only the frequency of vertices without introducing
forbidden vertices. But if the degree of randomization has reached a
certain level the number of forbidden vertices starts to rise rapidly.
The rhombohedron tiling contains dodecahedra enclosing two prolate and
two oblate rhombohedra. The internal vertices are
called simpletons and are arranged in two-dimensional
layers perpendicular to two-fold symmetry axis. A spin can be assigned
to the simpletons\cite{dotera94} depending on their position. 
In the ideal tiling all the spins in a certain layer carry spin +1 or --1.
With a proper summation rule a sheet magnetization $M$ of value 1 is formed.
In a random tiling the sheets exist
with a reduced value since the spins are no longer aligned.  
The susceptibility is given by 
$\chi=1/T<\!N_D>(<M^2/N^2_D>-<M/N_D>^2)$,
$N_D$ is the number of spins.

\section{Energy measures}
\label{energy}

In this work we deal with canonical rt ensembles. All
configurations of one ensemble have the same volume and the same
number of particles. The pure entropic rt model, on the
other hand, is specialized for microcanonical ensembles, since all
configurations have the same energy. Starting from a canonical rt
ensemble in the thermodynamic equilibrium we can calculate the 
internal energy as the ensemble average  $U=\left<E\right>$ of the
instantaneous
energy $E$. The specific heat $C_{V}(T)$ may then be derived from the
variance
$<(\delta E)^2>$. %= k_B T^2 C_V$. 
The entropy density $s(T)$~\footnote{Large letters indicate total
quantities, small letters denote quantities per tile.} is given by the
integration of the specific heat $c_V$.
The ground 
state entropy $s_0$ represents the number of states
energetically equivalent to the quasiperiodic ground state.
At $T\rightarrow\infty$ we
are in the limit of the pure rt model and we get the
configuration entropy $s_{\infty}$. The temperature
variation of the $c_{V}$ depends 
on the energy measure, but $s_{\infty}$ is independent of it.
%If one is interested only in $s_{\infty}$, then the specific choice of
%the energy measure has no physical relevance provided the following
%conditions are fulfilled: 
This is especially the case if the following holds:
First, the energy of any configuration is unique.
Second, the quasiperiodic reference tiling is a ground state.
For global energy measures states with an energy less than the
quasiperiodic reference tiling may exist which would indicate that
this state is not stable at $T=0$.
Third, the energy measure is limited from above. 
This is a requirement for the integrability of
$c_V(T)/T$ as a function of $T$. If the energy measure is not
limited, uncontrolled fluctuations of the energy in the
high-temperature limit may exist. If they vary stronger than quadratic
with $T$ then $<(\delta E)^2>/T^2$ diverges together with $c_V$ as
$T$$\rightarrow$$\infty$.

A globally defined energy measure with easily calculable ground state
entropy is the ``harmonic energy measure''.
The energy is given by the sum of the squared distances of the dual
quasilattice sites from their center of mass. Up to a factor $N$ it is
equal to the variance $\Omega$ in the orthogonal space.
The energy of the configuration $\alpha$ is given by
$E^{(\alpha)}=C N | \Omega^{(\alpha)} - \Omega^{(0)} |$~(2).
The index 0 denotes the ideal reference configuration,
$C$ is a normalization constant.
The variance for the ideal reference configuration 0 is smaller than
the variances for the overwhelming majority of the rt
configurations $\alpha$. But there is a tiny minority of
configurations with a variance smaller than the value of the ideal
tiling. Their atomic hypersurfaces are closer to a sphere than the
triacontahedron. 
To avoid
energies less than the energy of the ideal tiling we have taken
the absolute value in Eq.2. For zero global phason
strain this energy measure is not degenerate. But
for periodic approximants the $N$ possibilities to
chose the origin of the unit cell yield a ground state entropy
$s_0=\ln{N}/N$ per lattice point which vanishes in the thermodynamic
limit. Strandburg~\cite{strandburg91} has
introduced a similar energy measure. But it was taken relative to a fixed
point in the orthogonal space and not relative to the center of mass and
therefore does not fulfil the criterium of finiteness of the energy
measure. Without fixed boundaries of the system
the whole distribution of vertex points in orthogonal space
may drift and therefore yield a systematic contribution to the energy.

\section{Results}
\label{results}

\begin{figure}
\caption{Specific heat $c$ and susceptibility $\chi$. Sizes are
indicated by the number of vertices. 
\label{specheat}\label{suscept}}
\centerline{
\hspace*{3mm}
\psfig{file=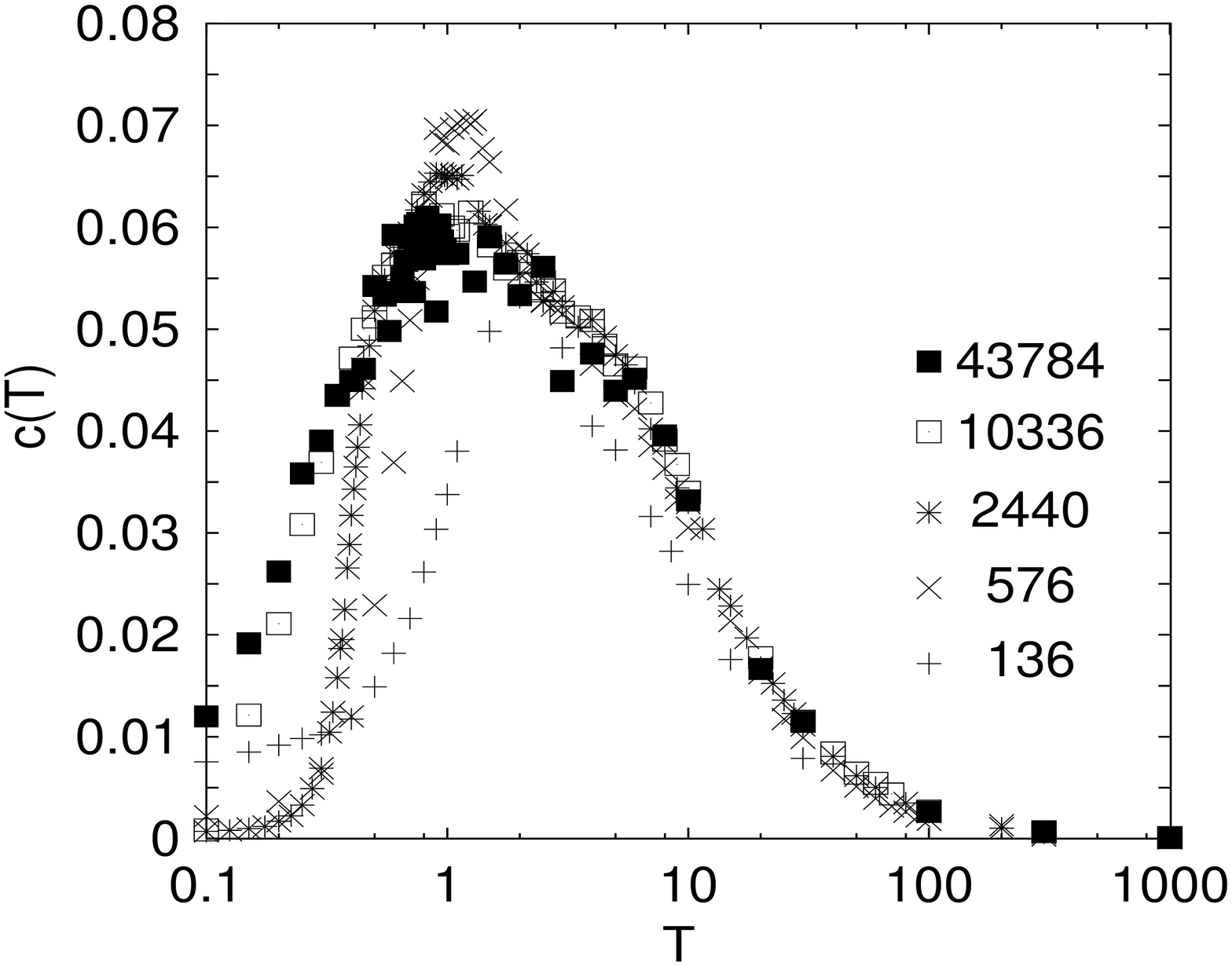,width=5.8cm,angle=270}
\hfill
\psfig{file=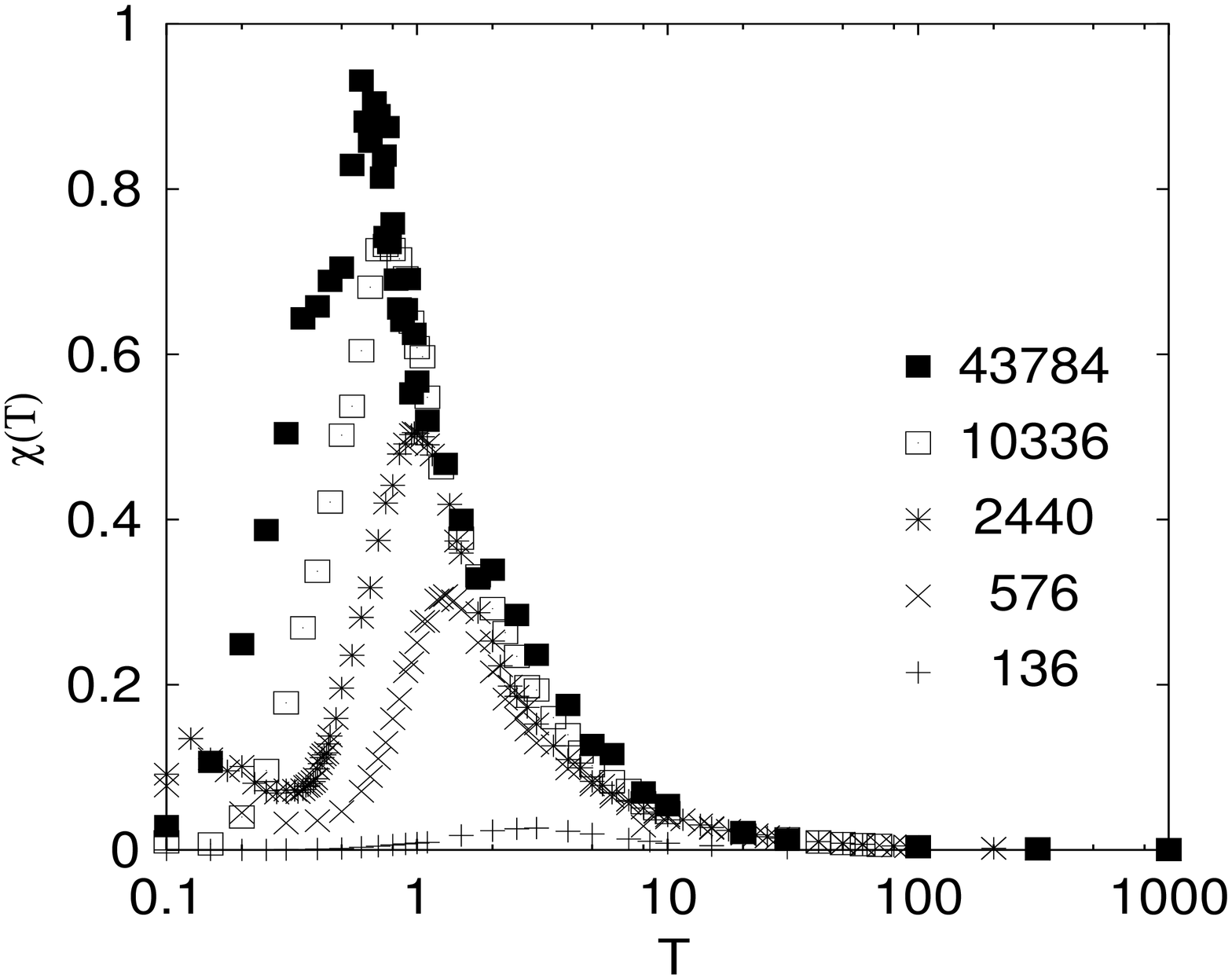,width=5.8cm,angle=270}
}
\end{figure}

A plot of the $c_V$ and $\chi$ vs.\ $T$ for different sample sizes
(finite-size scaling) may disclose a second order rt phase transition
if it diverges with sample size. To prove this, ensemble averages were
calculated for five cubic  approximants from 136 ($n=3$), up to
43784 ($n=7$) vertices ($n$ is the generation).  
The internal energy grows monotonously but saturates at $T \rightarrow
\infty$ at a value depending on the size of the
sample. The limit can be derived from limits of the variance: 
$\Omega(T=\infty) = 1.73 \pm 0.01$ and leads to $u$ between 1.97 and 2.05.
A \textsc{Schottky} anomaly is present in the $c_V$-plot
(Fig.~\ref{specheat}) but the specific heat shows an additional bump
above the maximum. Such a behaviour is known for few-level systems
with sufficiently separated levels.
We have mapped the distribution of the energy levels for $T=\infty$.
No indication of discrete energy levels was found, only an asymmetry
of the distribution with a smaller slope at higher energies was observed.
There exist other energy measures~\cite{gaehler95} which
exhibit no visible asymmetry in the energy distribution and no bump in
the specific heat. 
%This, however, in our opinion is not a clear explanation for the
%the shape of $c_{V}$ in this work. 
The value of the maximum of $c_{V}$ is not significantly
dependent on sample size.
The increase in $s_{\infty}$ is caused only
by the growing width of the maximum.
The reason why there is no divergence of the maximum of $c_V$
may be that the
intrinsic divergence of the specific heat with sample size, if any, is very
weak. For closer insight we calculated the sheet magnetization $M$ and the
susceptibility $\chi$ since the latter shows a much more
pronounced divergence behaviour~\cite{dotera94}.
The value of the maximum of the susceptibility
(Fig.~\ref{suscept}) grows about linearly with the generation $n$ and
moves to lower $T$. It is not yet clear if the relation $\chi_{max}
(n) \propto (n-n_0)$ is valid for $n > 7$. If yes, this would be a
slow divergence (more precisely: $\chi_{max}$($N$) is about
proportional to $N^{4.25}$, thus indicating a transition at $T=0$.
The alternation condition~\cite{dotera94} yields much
clearer results for $c_{V}$ and $\chi$ --- maybe as a consequence of
its closer similarity to an Ising model.

The mean square displacement grows
linearly with time $t$, indicating a normal diffusion behaviour.
The diffusion coefficient forms a plateau at $T>1$ for
$n=4,5,6$ in the \textsc{Arrhenius} plot. There may be several reasons
for this behaviour:
First, there are energy barriers which at low temperatures
lower the mobility of lattice points for higher energy flips.
In the range of the plateau the probability for a flip only occasionally
suffices to overcome the barriers which play no role at high temperatures.
Second, 
a phase transition may exist which changes the slope in the \textsc{Arrhenius}
plot. This is how G\"ahler explained a similar
behaviour for the alternation condition.
At last, the number of flipable lattice points may change with temperature. 
The number of simpletons is about 23\% in the range $0< T< 1$. It
decreases up to $T\approx 10$. The plateau of $D(T)$ is most clearly
seen in this range. Up to $T \approx 100$ the further decline leads
to an increase of 
the negative slope of $D(T)$. Above $T \approx 100$ the number of
simpletons is constant at $\approx 17.5\%$. 
The behaviour of $D(T)$ is obscured to some degree by 
modes with zero energy cost caused by periodic boundaries. These
modes become less important at larger sizes, but they suppress
the plateau for small sample~sizes.

\end{document}